\documentclass[prb,twocolumn,showpacs,aps]{revtex4}
\usepackage{graphicx}

\begin{document}

\title{Extrinsic photonic crystals}
\author{Chun Xu, Dezhuan Han, Xin Wang, Xiaohan Liu,}
\author{Jian Zi}
\email{jzi@fudan.edu.cn}
\affiliation{Surface Physics Laboratory and Department of Physics, Fudan University,
Shanghai 200433, People's Republic of China}
\date{\today }

\begin{abstract}
Doped semiconductors are intrinsically homogeneous media. However, by
applying an external magnetic field that has a spatially periodic variation,
doped semiconductors can behave extrinsically like conventional photonic
crystals. We show this possibility theoretically by calculating the photonic
band structures of a doped semiconductor under an external, spatially
periodic magnetic field. Homogeneous media, behaving like conventional
photonic crystals under some external, spatially periodic fields, define a
new kind of photonic crystals: extrinsic photonic crystals. The proposed
extrinsic photonic crystals could not only extend the concept of photonic
crystals but also lead to the control of the dispersion and propagation of
electromagnetic waves in a unique way: simply manipulating the externally
applied fields.
\end{abstract}

\maketitle

In the areas of computing and communication there has been a strong desire
to replace electronic devices with photonic ones due to the fact that as
information carriers electromagnetic waves are advantageous in many ways
over electrons. One of the promising approaches is based on photonic
crystals (PCs).\cite{yab:87,joh:87} PCs proposed up to now are composite
materials with a permittivity or/and a permeability which is a periodic
function of the position.\cite{joa:95} As a result of the multiple Bragg
scatterings, PCs are characterized by complicated photonic band structures.
Between photonic bands there may exist photonic band gaps (PBGs), for
frequencies within which the propagation of electromagnetic waves is
absolutely forbidden. The existence of the complicated photonic band
structures and PBGs in PCs allows the control of dispersion and propagation
of electromagnetic waves somewhat in a desired way, which can lead to many
novel applications.\cite{sou:96,sak:01,ino:04,lou:05}

In conventional PCs the spatially periodic variation of the permittivity
or/and the permeability is obtained by the periodic arrangement of two or
more materials. To obtain more degrees of tunability, tunable PCs have been
proposed. The tunability relies on the modification of the permittivity
or/and the permeability of the constituent materials by some external
parameters such as temperature, external electric or magnetic fields.\cite%
{fig:98,bus:99,kee:00,leo:00,hal:00,kee:01,li:03,jia:03,xu:03,tian:05}
Tunable PCs proposed up to now still consist of two or more materials. Based
on tunable PCs, it is possible to design and fabricate new kinds of
optoelectronic and microwave devices such as optical modulators, switches,
tunable filters, and tunable resonators.

In the present work we propose and conceptualize a new kind of tunable PCs:
extrinsic PCs. Unlike conventional PCs, an extrinsic PC is composed of a 
\textit{single} homogenous material, whose permittivity or/and permeability
can be altered by applying some external fields. If the applied external
filed is spatially periodic, a spatially periodic variation of the
permittivity or/and the permeability can be obtained likewise. Consequently,
this homogenous material behaves like conventional PCs. Extrinsic PCs may
extend the concept of PCs, leading likely to some new applications. To
exemplify the idea of extrinsic PCs, we present theoretical calculations of
the photonic band structures for a doped semiconductor under a spatially
periodic magnetic field. Our results indicate that this doped semiconductor
behaves extrinsically like a conventional PC.

In previous works\cite{jia:03,xu:03,tian:05} we showed that PCs consisting
of doped semiconductors can be made tunable under an external magnetic
field. The central idea relies on the fact that the dielectric constant of
doped semiconductors can be altered by applying an external magnetic field
owing to magneto-optical effects.\cite{per:67,pid:80} In the present work,
we take advantage of one of the famous magneto-optical effects, Voigt effect,%
\cite{per:67,pid:80} in order to achieve extrinsic PCs. The proposed
extrinsic PCs here are composed of a single $n$-doped semiconductor. For
frequencies well below the phonon resonance frequency, the dielectric
constant of $n$-doped semiconductors is given, in the absence of the
external magnetic field, by\cite{kit:76} 
\begin{equation}
\varepsilon (\omega )=\varepsilon _{0}\left( 1-\frac{\omega _{p}^{2}}{\omega
^{2}}\right) ,  \label{de}
\end{equation}%
where $\varepsilon _{0}$ is the static dielectric constant. The plasma
frequency $\omega _{p}$ is obtained from 
\begin{equation}
\omega _{p}^{2}=\frac{4\pi ne^{2}}{m^{\ast }\varepsilon _{0}},
\end{equation}%
where $n$ is the density of electrons, $e$ is the effective charge of
electrons, and $m^{\ast }$ is the effective mass of electrons. In Voigt
configuration,\cite{per:67,pid:80} the propagation direction of
electromagnetic waves is perpendicular to the applied magnetic field. The
modification of the dielectric constant due to the external magnetic field
is different for different polarizations. For $E$-polarization (with the
electric field parallel to the external magnetic filed), the dielectric
constant is not affected by the applied magnetic field, still given by Eq. (%
\ref{de}). For $H$-polarization (with the electric field perpendicular to
the external magnetic field), however, the dielectric constant is modified
in the presence of the external magnetic field, given by\cite{pid:80} 
\begin{equation}
\varepsilon (\omega )=\varepsilon _{0}\left( 1-\frac{\omega _{p}^{2}}{\omega
^{2}-\omega _{c}^{2}}-\frac{\omega _{p}^{4}\omega _{c}^{2}}{\omega
^{2}\left( \omega ^{2}-\omega _{c}^{2}\right) \left( \omega ^{2}-\omega
_{c}^{2}-\omega _{p}^{2}\right) }\right) .  \label{dh}
\end{equation}%
The cyclotron frequency is $\omega _{c}=eB/m^{\ast }c$, where $B$ is the
amplitude of the external magnetic field and $c$ is the speed of light in
vacuum. It should be noted that for frequencies substantially above the
phonon resonance, $\varepsilon _{0}$ in Eqs. (\ref{de}) and (\ref{dh})
should be replaced by the optical dielectric constant $\varepsilon _{\infty
} $. It is obvious that for $H$-polarization the dielectric constant is a
function of the external magnetic field $B$. Thus, a spatially periodic
variation of the refractive index can be achieved provided that the applied
external magnetic field is spatially periodic.

\begin{figure}[thpb]
\centerline{\includegraphics[angle=0,width=7cm]{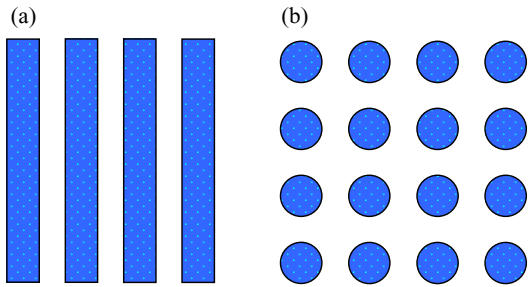}}
\caption{(color online). Schematic view of the externally applied magnetic
field with (a) 1D and (b) 2D spatially periodic variations. The magnetic
field is perpendicularly applied only to the shaded regions.}
\label{fig1}
\end{figure}

We now consider an external magnetic field that has one-dimensional (1D) and
2D spatially periodic variations as shown schematically in Fig. \ref{fig1}.
As a first approximation, it is assumed that the magnitude of the magnetic
field is homogeneous everywhere in the applied region, while it is zero
outside the applied region. In real cases, the applied magnetic field is not
so ideal as assumed, which may lead to some quantitative changes in our
results. But it does not affect our conclusions. Under the spatially
periodic magnetic field, a doped semiconductor should behave like a
conventional PC. Without loss of generality, the doped semiconductor is
assumed to be $n$-doped GaAs. The static dielectric constant of GaAs is $%
\varepsilon _{0}=12.9$, taken from Ref. \onlinecite{kit:76}.

\begin{figure}[thpb]
\centerline{\includegraphics[angle=0,width=8cm]{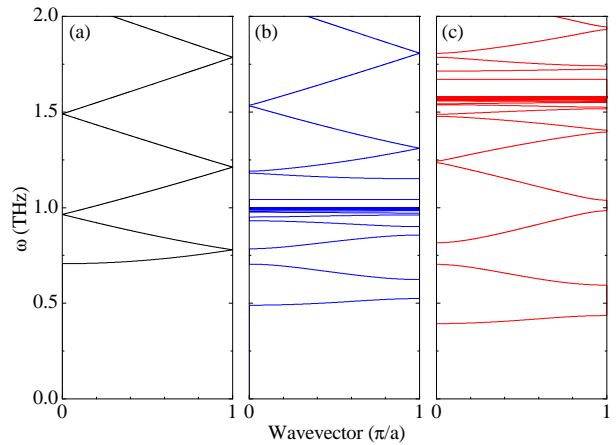}}
\caption{(color online). Calculated photonic band structures of $H$%
-polarization for $n$-doped GaAs with $\protect\omega _{p}=0.707$ THz under
a 1D spatially periodic magnetic field. The magnitude of the applied
magnetic field is (a) 0, (b) 0.264, and (c) 0.528 Tesla.}
\label{fig2}
\end{figure}

With a 1D spatially periodic magnetic field applied to $n$-doped GaAs as
shown in Fig. \ref{fig1}(a), $n$-doped GaAs behaves like a 1D PC, whose
photonic band structures for $H$-polarization as a function of the magnitude
of the applied magnetic field are shown in Fig. \ref{fig2}. The photonic
band structures are calculated from a transfer matrix method.\cite%
{yeh:88,zi:98} In photonic band structure calculations, the lattice constant
of the periodic magnetic field is $a=0.8$ mm and the fraction of the applied
region with respect to one unit cell is $0.7a$.

For $E$-polarization, the photonic band structures are not affected by the
applied magnetic field, which are exactly the same as those of $H$%
-polarization in the absence of the magnetic field. There exists a
low-frequency band gap that extends up to the plasma frequency $\omega _{p}$%
. Above $\omega _{p}$, the photonic band structure is simply a folded
version from the corresponding dispersion of $n$-doped GaAs. With the
applied magnetic field, drastic changes in the photonic band structures for $%
H$-polarization occur. The upper edge of the low-frequency band gap shifts
downwards with the increasing magnetic field. Below $\omega _{p}$,
additional photonic bands and PBGs appear. Above $\omega _{p}$, some PBGs
open up owing to the multiple Bragg scatterings. Just below $(\omega
_{p}^{2}+\omega _{c}^{2})^{1/2}$, there exist very dense, flat photonic
bands owing to the fact that the region applied with the magnetic field has
a very large positive dielectric constant. For frequency well above $\omega
_{p}$, photonic band structures are less affected due to the fact that the
dielectric constant is weakly modified at high frequencies.

\begin{figure}[thpb]
\centerline{\includegraphics[angle=0,width=8cm]{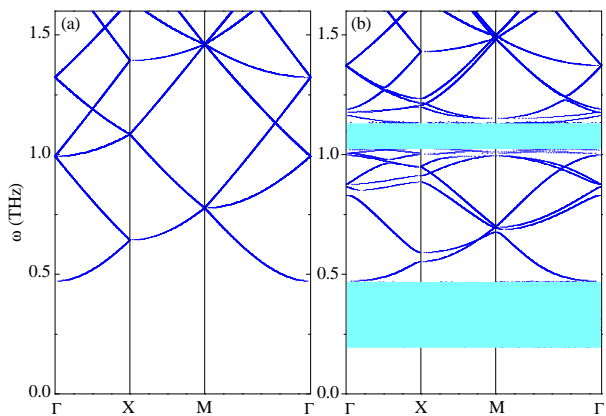}}
\caption{(color online). Calculated photonic band structures of $H$%
-polarization for $n$-doped GaAs with $\protect\omega _{p}=0.471$ THz under
a 2D spatially periodic magnetic field. The magnitude of the applied
magnetic field is (a) 0 and (b) 0.353 Tesla. The high symmetrical points in
the irreducible Brillouin zone of a square lattice are denoted by $\Gamma
=(0,0)$, $X=(1,0)\protect\pi /a$, and $M=(1,1)\protect\pi /a$. In the shaded
regions there appear very dense, flat photonic bands.}
\label{fig3}
\end{figure}

When a 2D spatially periodic magnetic field is applied to $n$-doped GaAs as
shown in Fig. \ref{fig1}(b), this $n$-doped GaAs should behave like a 2D PC.
The lattice type of the periodically applied magnetic field is square with a
lattice constant of $a=0.6$ mm. The applied regions have circular shapes
with a radius of $0.4a$. A plane-wave-based transfer matrix method\cite%
{li:03b} is used to calculate the photonic band structures, shown in Fig. %
\ref{fig3}. Without the external magnetic field, the photonic band
structures for $E$- and $H$-polarizations are degenerate. A band gap exists
at the frequencies below $\omega _{p}$ and there are no PBGs above $\omega
_{p}$. The photonic band structures are a simple folding from the dispersion
of $n$-doped GaAs. When a spatially periodic magnetic field is applied, some
changes occur in the photonic band structures for $H$-polarization. Unlike
in 1D systems, no complete PBGs open up, owing to the fact that the
modification by the applied magnetic field is not strong enough to give rise
to a strong contrast of the refractive index. However, some additional
photonic bands and partial PBGs appear. For example, at $B=0.353$ Tesla
there exists a partial PBG ranging from 0.555 to 0.587 THz along the $\Gamma
X$ direction. Similar to the 2D PC composed of a doped semiconductor
perforated with air hole arrays,\cite{xu:03} some very dense, flat photonic
bands are present in two frequency ranges, one from 1.052 to 1.137 THz and
other one from 0.195 to 0.471 THz. The existence of the dense, flat photonic
bands (shaded regions) is due to fact that the dielectric constant of one of
the regions applied with and without the magnetic field is negative, while
the dielectric constant in other region is positive. The low-frequency PBG
still exists for frequencies below 0.195 THz, where the dielectric constants
of the two regions are both negative. For frequencies well above $\omega
_{p} $, the photonic band structures are weakly affected as expected.

\begin{figure}[thpb]
\centerline{\includegraphics[angle=0,width=7.5cm]{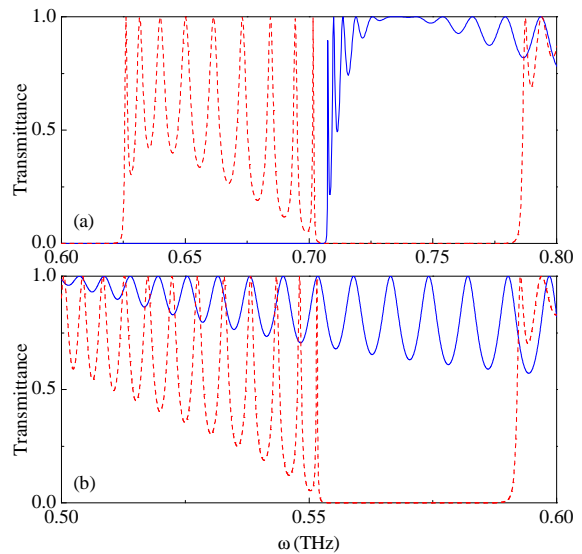}}
\caption{(online color). Transmittance spectra of $H$-polarization for
finite $n$-doped GaAs under (a) 1D and (b) 2D spatially periodic magnetic
fields. The thickness of GaAs is 8 mm in 1D and 19.2 mm in 2D, respectively.
The other structural parameters for 1D and 2D are the same as used in Figs. 
\protect\ref{fig2} and \protect\ref{fig3}. The magnitude of the applied
magnetic field is 0.264 and 0.353 Tesla for 1D and 2D, respectively. Solid
(dashed) lines stand for transmission in the absence (presence) of the
magnetic field. }
\label{fig4}
\end{figure}

From the above discussions, we show the possibility to achieve extrinsic PCs
based on a single $n$-doped semiconductor. It is know that the plasma
frequencies of $n$-doped semiconductors lie in the THz regime for reasonable
doping densities. Therefore, the proposed GaAs-based extrinsic PCs may have
some potential applications in the THz technology such as switches,
modulators, tunable filters and resonators. As an example, we show in Fig. %
\ref{fig4} the switching effects of finite $n$-doped GaAs under 1D or 2D
spatially periodic magnetic filed. For the 1D case, for wavelengths in the
vicinity of 0.75 THz, electromagnetic waves can transmit in the absence of
the applied magnetic filed. When the magnetic filed is applied, there is no
transmission due to the existence of a PBG. For wavelengths in the vicinity
of 0.66 THz, there is transmission when the magnetic field is applied, while
electromagnetic waves cannot transmit when the magnetic filed is switched
off. Similarly, this switching occurs also in the 2D system.

It should be pointed out that point and line defects can be also introduced
in extrinsic PCs. This can be achieved by introducing point and line defects
in the external, spatially periodic magnetic field. Consequently, tunable
cavity and waveguides could be obtained. In the above discussions, the
external magnetic field is static. However, the applied magnetic field can
be also alternative. As a result, a dynamic control of the optical
properties of extrinsic PCs can be achieved, which may lead to some novel
applications.

In summary, we proposed and conceptualized a new kind of tunable PCs:
extrinsic PCs. Without externally applied fields, they are homogeneous
media; when external fields are applied, they are PCs simultaneously. We
showed by numerical calculations that a $n$-doped GaAs behaves like
conventional PCs under a spatially periodic magnetic field. Our
conceptualized extrinsic PC may greatly extend the concept of PCs. Moreover,
by manipulating the externally applied fields extrinsic PCs provide with a
unique way to control the dispersion and propagation of electromagnetic
waves.

This work was mostly supported by CNKBRSF. Partial support from NSFC,
PCSIRT, and Shanghai Science and Technology Commission is also acknowledged.
One of the authors (J.Z.) thanks Prof. Yanfeng Chen for interesting
discussions.

\end{document}